\documentclass[aps,prd,twocolumn,groupedaddress,nofootinbib]{revtex4}

\usepackage{amsfonts,amsmath,amssymb}
\usepackage{mathrsfs}
\usepackage{nicefrac}
\usepackage{graphicx}
\usepackage{color}
\usepackage{soul} 
\usepackage{hyperref}
\usepackage{accents}
\usepackage{subfigure}
\usepackage[normalem]{ulem}
\usepackage{cleveref}
\hypersetup{
    colorlinks=true, 
    linktoc=page,    
    linkcolor=blue,  
    urlcolor=black
}

\newcommand{\be}{\nopagebreak[3]\begin{equation}}
\newcommand{\ee}{\end{equation}}

\newcommand{\bee}{\nopagebreak[3]\begin{equation*}}
\newcommand{\eee}{\end{equation*}}

\newcommand{\ba}{\nopagebreak[3]\begin{eqnarray}}
\newcommand{\ea}{\end{eqnarray}}

\newcommand{\baa}{\nopagebreak[3]\begin{eqnarray*}}
\newcommand{\eaa}{\end{eqnarray*}}

\newcommand{\la}{\label}
\newcommand{\n}{\nonumber}

\begin{document}
\title{
The black hole quantum atmosphere}

\author{Ramit Dey}
\email[]{rdey@sissa.it}

\author{Stefano Liberati}
\email[]{liberati@sissa.it}

\author{Daniele Pranzetti}
\email[]{dpranzetti@sissa.it}

\affiliation{SISSA, 
Via Bonomea 265, 34136 Trieste, Italy and INFN, Sezione di Trieste}
                
\begin{abstract}
Ever since the discovery of black hole evaporation, the region of origin of the radiated quanta has been a topic of debate. Recently it was argued by Giddings that the Hawking quanta originate from a region well outside the black hole horizon by calculating the effective radius of a radiating body via the Stefan--Boltzmann law. In this paper we try to further explore this issue and end up corroborating this claim, using both a heuristic argument and a detailed study of the stress energy tensor. We show that the Hawking quanta originate from what might be called a quantum atmosphere around the black hole with energy density and fluxes of particles peaked at about $4MG$, running contrary to the popular belief that  these originate from the ultra high energy excitations very close to the horizon. This long distance origin of Hawking radiation could have a profound impact on our understanding of the information and transplanckian problems.
\end{abstract}

\maketitle
\section{Introduction}
The discovery of Hawking radiation~\cite{hawking1975} changed our perspective towards black holes, giving us a deeper insight about the microscopic nature of gravity. At the same time, within the semi-classical framework, the current understanding of such process still leaves open several issues. Of course, a well known unresolved problem of black hole physics is the information loss paradox~\cite{hawking1982, Page:1993wv, Giddings:1992hh}, i.e.~the apparent incompatibility between the complete thermal evaporation of a black hole endowed with an event horizon and unitary evolution as prescribed by quantum mechanics. 

For restoring unitarity of Hawking radiation and addressing the information loss problem correctly, it is important (among other things) to know from where the Hawking quanta originate. For example, if one assumes a near horizon origin of the Hawking radiation, then one way to restore unitarity is by conjecturing some sort of UV-dependent entanglement between partner Hawking quanta which would enable the late time Hawking flux to retrive the information  in the early stages of the evaporation process. Such scenario seems to lead to the so called  ``firewall'' argument as the conjectured lack of maximal entanglement between the Hawking pairs makes the near horizon state singular and eventually demands some drastic modification of the near horizon geometry~\cite{Almheiri:2012rt}. On the other hand, if one believes in a longer distance origin of the Hawking quanta, some effect must be operational at a larger scale for restoring unitarity rather than near the horizon, avoiding the ``firewall''. 

A similar open issue is the transplanckian origin of Hawking quanta. Hawking's original calculation indicates that the quanta originate near the black hole horizon in a highly blue-shifted state requiring an assumption on the UV completion of the effective field theory used for the computation and on the lack of back-reaction on the underlying geometry \footnote{See, for instance, \cite{Pranzetti:2012pd, Pranzetti:2012dd} for a black hole evaporation analysis where these issues can be addressed in a quantum gravity context.}. While it was debated for a while if Hawking quanta could originate initially, during the star collapse, and later released over a very long time, it was convincingly argued in~\cite{PhysRevD.15.365} that this cannot be the case if an event horizon indeed forms. This leads to the conclusion that the Hawking quanta are generated in a region outside the horizon. 
A conclusion corroborated by studies of the Hawking modes correlation structure where it was shown that mode conversion happens over a long distance from the horizon~\cite{Parentani:2010bn}.
A more recent claim in this direction, based on calculating the size of the radiating body via the Stefan--Boltzmann law, showed that the Hawking quanta originate in a near horizon quantum region, a sort of black hole ``{\it atmosphere}"~\cite{Giddings:2015uzr}. It is a well known fact that the typical wavelength of the radiated quanta is comparable to the size of the black hole, so one might think that the point particle description is not very accurate. However, as measured by a local observer near the horizon, the wavelength is highly blue-shifted when traced back from infinity to the horizon, thus validating the point particle description. 

 The Hawking process can be explained  heuristically as-well, for example via a tunnelling mechanism where the particle tunnels out of the horizon or the anti particle (propagating backwards in time) tunnels into the horizon and as a result of this we get the constant Hawking flux at infinity \cite{Parikh:1999mf}. Alternatively, 
one popular picture is to imagine that the strong tidal force near the black hole horizon stops the annihilation of the particle and anti-particle pairs that are formed spontaneously from the vacuum. Once the antiparticle is ``hidden" within the black hole horizon, having a negative energy effectively, the other particle can materialise  and escape to infinity \cite{hawking1979general,Adler:2001vs}. 

In this paper we shall explicitly make use of this latter heuristic picture as well as of a full calculation of the stress energy tensor in 1+1 dimensions. We shall see that both  methods seem to agree in suggesting that the Hawking quanta originate from the black hole {\it atmosphere} and not from a region very close to the horizon. In section II, based on the heuristic picture of Hawking radiation described above and invoking the uncertainty principle and tidal forces, we show that most of the contribution to the radiation spectrum comes from a region far away from the horizon. In section III we further strengthen our claim by a detailed calculation of the renormalized stress energy tensor, which indicates a similar result.

\section{A gravitational Schwinger effect  argument} \label{Schwinger}

One ingredient of our heuristic argument to identify a quantum atmosphere outside the black hole horizon, where particle creation takes place, is the uncertainty principle. However, the use of the uncertainty principle alone, as originally suggested by Parker~\cite{Parker1977}, does not contain any physically relevant information about the location of particle production and why smaller black holes should be hotter. 
Indeed, the uncertainty principle in this case  provides a rough estimate of the region of particle production as inversely proportional to the energy of the Hawking quanta when they are produced, but it does not take into account any dynamical mechanism to estimate  the probability of spontaneous emission.

Thus one can improve this argument by invoking a physical process of creation of the Hawking quanta and using the uncertainty principle as a complementary tool to estimate the region of origin of the quanta. In this section, we try to achieve this goal by relying on tidal forces.

 Let us then consider a situation where a virtual pair, consisting of a particle and anti-particle, pops out of the vacuum spontaneously for a very short time interval  and then annihilates itself. 
In the Schwinger effect~\cite{Schwinger:1951nm} a static electric field is assumed to act on a virtual electron-positron pair until the two partners are torn apart once  the threshold energy necessary to become a real electron-positron pair is provided by the field. Energy is conserved due to the fact that the electric potential energy has opposite sign for partners with opposite charge. However, in its gravitational counterpart a priori only vacuum polarisation can be induced by a static field in the absence of an horizon.

In fact, only in the presence of the latter one has both the characteristic peeling structure of geodesics (diverging away from the horizon on both its sides) as well as the presence of an ergoregion behind it.~\footnote{This is strictly true only for non-rotating black holes, for rotating ones the ergoregion lies outside of the horizon allowing for the classical phenomenon of superradiance. However, the quantum emission still requires the peculiar peeling structure of geodesics typical of the horizon.} The presence of an ergoregion is crucial for energy conservation as it allows for negative energy states given that in it the norm of the timelike Killing vector, with respect to which we compute energy, changes sign.

Indeed, if a Schwinger-like process takes place near the black hole horizon, due to the tidal force of the black hole and the peeling of geodesics, the pair can get spatially separated  and one partner can enter the black hole horizon following a timelike or null curve with negative energy while the other particle can escape to infinity and contribute to the Hawking flux.  In this picture, we are implicitly assuming that virtual particles in the vicinity of a black hole horizon move along geodesics when they are just about to go on-shell. 

Therefore, the physical scenario we want to envisage is that of a particle-antiparticle pair pulled apart by the black hole tidal force outside the horizon until they go on-shell as one of them reaches the horizon \footnote{One could also consider the case where the ingoing particle tunnels through the horizon and goes on-shell well inside the horizon (as e.g.~suggested by the results of~\cite{Parentani:2010bn}); however, since in our analysis below we are interested in the tidal force as computed in the outgoing particle rest frame, this should not affect the final expression for the force. Thus, from the point of view of an outside static observer, the work done by the gravitational field on the pair (in our heuristic derivation) is insensitive to the exact location where the ingoing particle becomes real.} located at $r_s=2GM/c^2$ (actually an infinitesimal distance inside it so that the geodesic motion will drag it further inside) while the other particle is at a radial coordinate distance $r=r_*$.  Once on-shell, the outgoing particle eventually reaches infinity and contributes to the Hawking spectrum. In order to do so though, it has to be created with an energy corresponding to the energy of the Hawking quanta at a distance $r_*> r_s$ from the center of the black hole as measured by a local static observer; this can be reconstructed by noticing that 
\be \la{omegarini}
\omega_r=\frac{\omega_\infty}{\sqrt{g_{00}}}\,,
\ee
where $\omega_\infty$ is the energy at infinity and we are using the $(+,-,-,-)$ signature. At infinity, the thermal spectrum of Hawking radiation gives
\be\la{omegaT}
\omega_\infty=\gamma \frac{k_B T_H}{\hbar}\,,
\ee
where the Hawking temperature for a black hole of mass $M$ reads 
$
k_B T_H=\frac{\hbar c^3}{8\pi GM}\,
$,
and  $\gamma$ is a numerical factor spanning the energy range of the quanta giving rise to the radiation thermal spectrum. At the peak of the spectrum $\gamma\approx 2.82$.

Thus,
we get
\be\la{omega}
\omega_\infty=\gamma\frac{c^3}{8\pi GM}
\ee
and
\be\la{omegar}
\omega_r
=\gamma\frac{c}{4\pi r_s}\frac{1}{\sqrt{1-\frac{\textstyle r_s}{\textstyle r}}}\,,
\ee

 This energy is provided by the work done by the gravitational field to pull the two partners apart. We can compute this work in the static frame outside a black hole and compare it with $\omega (r_*)$. Using this relation,  we can determine the region from which the Hawking quanta originate. This is the process we now want to implement. {Although in the rest of this Section we present the detailed derivation of the relation between the outgoing particle energy and the radial distance at which it goes on-shell for the massive case, our result holds also for massless particles. We comment at the end of this Section on how the same Schwinger effect argument can be implemented straightforwardly to the massless case.}

Let us clarify that, in a general relativistic framework, the geodesic deviation equation  does not describe the force acting on a particle moving along a geodesic. Rather, it expresses how the spacetime curvature influences two nearby geodesics, making them either diverge or converge, i.e.~it effectively measures tidal effects. Therefore, we can interpret these effects as the pull of the gravitational force on particles and talk about the work done by the gravitational field only in an heuristic sense. Nevertheless, in the case considered here where 
 the test particles have a mass much smaller than the black hole and we can neglect back-reaction effects, we expect this interpretation of the gravitational field effects to capture some relevant aspects of black hole physics. 
With these assumptions spelled out, let us proceed. 

In the rest frame of the outgoing particle,  one would see the antiparticle accelerating towards the horizon due to the tidal force. This radial acceleration in the rest frame of the particle can be computed using the geodesic deviation equation, namely
\ba\la{ar}
\left . a^r\right|_{r_*}\equiv \left . \frac{Dn^r}{D \tau^2}\right|_{r_*}=\left . {R^r}_{\mu\nu\rho}u^{\mu}u^{\nu}n^{\rho}\right|_{r_*}\,,
\ea
where the r.h.s. is expressed in terms of the Riemann tensor components,  $n^r$  denotes the separation between the two radially infalling geodesics followed by the pair of particles and $u^{\mu}=[1,0,0,0]$ in the rest frame of the particle. 

The separation between the particle and the anti-particle when the pair forms spontaneously (i.e. they go ``on-shell'') is given by their Compton wavelength, namely $n^{\rho}=[0,n^r,0,0]$ where $n^r \sim \lambda_C=\hbar/mc$, and $m\ll M$ is the particles rest mass (from now on we shall work in units where $\hbar=c=1$).
So in the end, Eq.~\eqref{ar} implies that the radial component of the tidal acceleration (as computed in the rest frame of the particle at coordinate $r_*$) is given by  \footnote{For computation of the acceleration in the rest frame of the particle we need the Riemann tensor in the inertial frame of the particle. One can compute the Riemann tensor in the static Schwarzschild coordinates and then boost it using the free-fall velocity of the particle as measured in the static frame. A feature of the Schwarzschild geometry is that the components of the Riemann tensor remains invariant under such a boost \cite{misner1973gravitation}. Thus, in \eqref{ar} we have $R_{rttr}=-2MG/r^3$.}

\be \label{a}
 a^r|_{r_*}=\frac{2M G}{r_*^3} \lambda_C
\ee

Our aim is to determine the work done on the spontaneously created particle pair  by the tidal force  in the static frame outside the black hole. For this we need to compute the tidal force as measured by a static observer outside the black hole at the instant when the outgoing partner goes on shell. This can be achieved by considering the particle rest frame and the static observer frame as locally two inertial frames: The latter sees the particle as moving with outward velocity given by the radial component of the geodesic tangent vector $u^r=dr/d\tau$. Once this is known,  we can derive the radial acceleration observed by the static observer by performing a boost with rapidity $\zeta=\tanh^{-1}({u^r})$.

We thus need to determine the instantaneous radial component of the free fall velocity of the outgoing particle when it goes on-shell. 
This can be computed from the geodesic equation and it is given by
\be
u^r=\frac{dr}{d \tau}=\sqrt{\frac{2MG}{r}\bigg( 1-\frac{r}{r_0}\bigg)}\,,
\ee
where $r_0$ comes as an integration constant corresponding to the coordinate distance at which the particle velocity goes to zero. Since we are interested in the value of the radial component of the geodesic tangent vector at the instant when the outgoing particle goes on-shell and becomes an Hawking quantum which eventually reaches infinity, we can take the integration constant $r_0 \rightarrow \infty$, i.e.~Hawking quanta can be created with zero velocity only at infinity.
Hence, we get
\be
\la{ur}
\left.u^r\right|_{r_*}=\sqrt{\frac{2MG}{r_*}}\,.
\ee

We can now boost the acceleration vector $a^\mu=(0,a^r,0,0)$, where $a^r$ given by \eqref{a}, with a velocity parameter given by \eqref{ur}, in order to determine 
the tidal force in the static frame $a^r_{\rm st}$. We get $a^r_{\rm st}=a^r \cosh({\zeta})=a_r(1-2MG/r)^{-1}$ so that
the radial component of the force under this transformation is given by 

\be\la{Ft}
\left.F^r_{\rm tidal-st}\right|_{r_*}=\left. \frac{m a^r_{\rm st}}{(1-2MG/r)} \right|_{r_*}\!\!=\frac{m\lambda_C}{(1-2MG/r_*)^2}\frac{2MG}{r_*^3} \,,
\ee
where we have rescaled the mass in the rest frame by the appropriate Lorentz factor, $(1-2MG/r_*)^{-1}$. 
Finally, using the fact that $\lambda_C\sim 1/m$, the magnitude of the force is given by
\be\la{Ft2}
||F^r_{\rm tidal-st}||= \frac{2MG}{r_*^3} \bigg( 1-\frac{r_s}{r_*}\bigg)^{-\frac{3}{2}}\,.
\ee

In analogy with the Schwinger effect,  we shall now assume that the work done by the tidal force to split the virtual pair  can be approximated by the product of the force computed above with the distance over which it appears to have acted, i.e.~the separation of the two Hawking quanta as they go on-shell as measured by a static observer at $r_*$. Given that we have assumed that the ingoing Hawking quantum goes on shell as soon as it can do so, i.e. at horizon crossing, this distance will coincide with the static observer's proper distance to the horizon $d(r_*)$.
 
Therefore,  the work required by the tidal force to split the pair apart is given by~\footnote{Alternatively, we could introduce a 4-vector $\ell^\mu=(0,\ell^r,0,0)$, with $||\ell||=\sqrt{g_{\mu\nu}\ell^\mu \ell^\nu}=d(r_*)$, and compute the work as $W_{\rm tidal}\sim \left. g_{rr}F^r_{\rm tidal-st} \ell^r\right|_{r_*}$. This would give the same result.}

\be
W_{\rm tidal}\sim ||F_{\rm tidal-st}^r|| \,d(r_*)=  \frac{2MG}{r_*^3}\bigg( 1-\frac{r_s}{r_*}\bigg)^{-\frac{3}{2}}d(r_*) \,,
\label{eq:work}
\ee
where $d(r_*)$ is given by
\ba \la{dr}
d(r_*)&=&\int_{r_s}^{r_*}\sqrt{g_{rr}}dr'\\
&=&\! r_s\!\left(\!\sqrt{\alpha(\alpha-1)}+\frac{1}{2}\log{\!\left[\alpha\! \left(1+\sqrt{1-\frac{1}{\alpha}}\right)^2\right]}\!\right)\,,\la{d}\n
\ea
and we have defined $\alpha\equiv r_*/r_s$.

 We can then equate this work to the total energy of the two Hawking quanta being created, namely $W_{\rm tidal}=2\omega_r$. 
 This  gives us
\ba \la{work}
 \frac{2MG}{r_*^3}\bigg( 1-\frac{2MG}{r_*}\bigg)^{-\frac{3}{2}}d(r_*)=\frac{\gamma}{2 \pi r_s}\bigg( 1-\frac{2MG}{r_*}\bigg)^{-\frac{1}{2}}\,.
\ea

 Finally,  from eq. (\ref{work}) we get
\ba \la{work2}
\gamma &=& \frac{2\pi}{\alpha^2}\left(1-\frac{1}{\alpha}\right)^{-\frac{1}{2}}\\
&\cdot&\!\!\left(\!1+\frac{1}{2\sqrt{\alpha^2-\alpha}}
\log{\!\left[\alpha\left(1\!+\!\sqrt{1-\frac{1}{\alpha}}\right)^2\right]}\right).\n
\ea

\begin{figure}[htb]
\includegraphics[scale=.8]{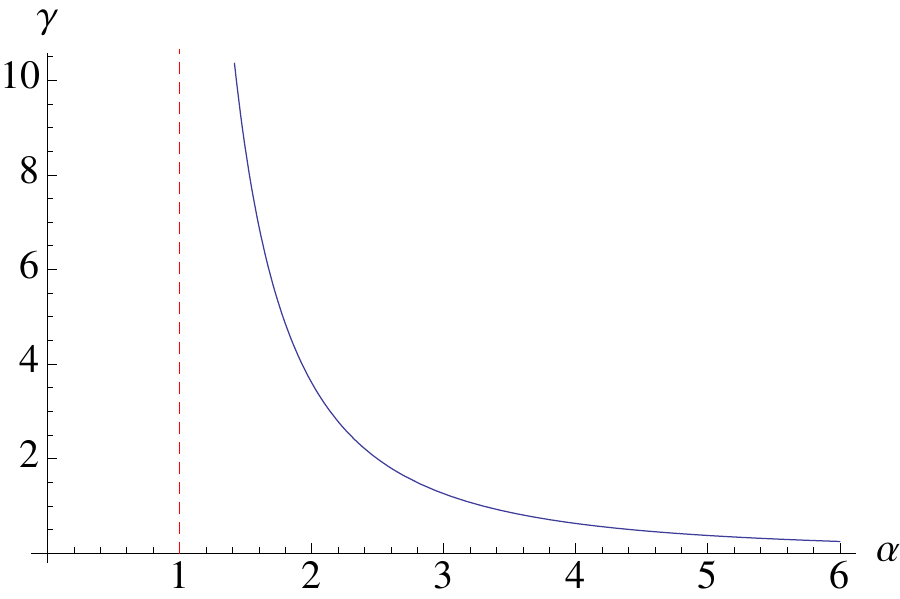}
\caption  {This plot shows the variation of $\gamma$ with respect to the radial distance from the center of the black hole. The red dashed line corresponds to the horizon location at $\alpha=1$ where the expression for the tidal force work diverges, indicating that the quanta in the far UV tail of the Hawking spectrum originate from very near the horizon.} \label{sigma-alpha}
\end{figure}

The relation between $\gamma$ and $\alpha$, i.e the radial distance scaled as $r_*/r_s$, is better illustrated in Fig. \ref{sigma-alpha}. It is clear from the plot that the part of the Hawking thermal spectrum around the peak $(\gamma \sim 2.82)$, where most of the radiation is concentrated, corresponds to a region which extends  far outside the horizon, up to around $ 2 r_s$ (at the peak $r_*\approx 4.38\, M G$). 

The plot above also shows how, in this tidal force derivation, the quanta with higher velocity (kinetic energy) are produced closer to the horizon. This is consistent with our analysis since the higher the initial radial velocity the stronger the Lorentz contraction of the outgoing particles distance from the horizon in their rest frame, given by $\lambda_C$, resulting in a shorter proper distance $d(r_*)$ at which they are detected.

Also, by using Eq.~\eqref{dr} and expressing the rest of Eq.~\eqref{eq:work} in terms of $\alpha$, we can see that the work doable at fixed $\alpha$ by the tidal forces  scales as the inverse of the mass of the black hole so making evident that smaller holes can produce hotter particles at the same relative distance from the horizon.

{ In the Schwinger effect argument we described in this Section we have considered the case of a massive test particle. However, in the physical context of a 4D Schwarzschild black hole, most of the radiation is emitted by massless particles. A generalization of our argument to the massless case can be achieved in a straightforward manner. In fact, 
despite the lack of a rest mass frame of one of the two partners, one can always study the Schwinger-like effect in a local inertial frame in the vicinity of the horizon and compute the radial acceleration \eqref{ar} considering  two radially infalling null geodesics with 4-velocity $u^{\mu}=[1,1,0,0]$ in such given frame; due to the symmetries of the Riemann tensor, this leads to the same expression \eqref{a} but with the Compton wavelength $\lambda_C$ replaced by the massless particle de Broglie wavelength $\lambda_B$. The acceleration as measured by a static observer outside the black hole then proceeds along the same lines as in the massive case, since the boost between the two frames, locally both inertial, is the same as in the massive case; we can thus compute the radial component of the tidal force as in \eqref{Ft}, where on the r.h.s. we replace the combination $m \lambda_C$ with $E \lambda_B$, $E$ being the massless particle energy, as measured by the static observer, which is related to the de Broglie wavelength by the standard relation $E=1/\lambda_B$ (recall that we switched to units where $\hbar=c=1$). In this way, we recover the expression \eqref{Ft2}, which is independent of the test particle mass. Therefore, the plot in FIG. \ref{sigma-alpha} applies also to the case of radiation being emitted by massless particles.
}

Let us stress again the heuristic nature of our argument. We are considering the instantaneous value of the tidal force observed by the outgoing partner at a given coordinate distance $r_*$ where it goes on-shell. However, we then use this instantaneous value to compute the work done by the gravitational field over a distance $d(r_*)$, as if the force was actually at work with the same constant value throughout the whole splitting process.  A similar approach was also used in \cite{grib1994vacuum} to give an estimate of the wavelength of the Hawking quanta as produced by the gravitational tidal force.

So, although the analogy with the Schwinger effect for the electron-positron pair production by an electric field may be advocated to lend support to our description of Hawking quanta production from a quantum atmosphere that extends well beyond the horizon, we  now want to  present  a more sound analysis based on the renormalized stress energy tensor in order to confirm this picture.

\section{Stress-energy tensor}\la{SET}
By analyzing the renormalized stress energy tensor (RSET) in the 2-dimensional case, one can understand Hawking radiation in a better way as this is a  local object which can help to probe the physics in the vicinity of the black hole. The derivation of the RSET components has been considered in many places in the literature \cite{PhysRevD.13.2720,Unruh:1977ga,birrell1984quantum,Singh:2014paa,Chakraborty:2015nwa}, here we build on these previous results and compute the energy density and flux as seen by an observer which has  zero radial velocity(thus giving rise to no kinematical effects) and zero acceleration at the horizon. 

\subsection{Computation of RSET}
Following~\cite{Barcelo:2007yk}, let us introduce a set of globally defined affine coordinates $U, V$ on $\mathscr{I}^-_{\rm left}, \mathscr{I}^-_{\rm right}$ respectively. Restricting to the radial and time dimensions, the metric reads
\be
ds^2=C(U,V) dU dV\,.
\ee
In $(1+1)$ dimensions the renormalised stress energy tensor for any massless scalar field  in terms of these affine null coordinates can be easily computed using the conformal anomaly \cite{PhysRevD.13.2720, Unruh:1977ga, Candelas:1980zt, birrell1984quantum,Padmanabhan:2003gd}. The components of the RSET computed in some arbitrary vacuum state are given as:
\ba
\langle T_{UU} \rangle&=&-\frac{1}{12\pi}C^{1/2}\partial^2_UC^{-1/2}
\n\\
&=&\frac{1}{24 \pi} \bigg[\frac{C_{,UU}}{C}-\frac{3}{2}\frac{(C_{,U})^2}{C^2} \bigg]\,,
\\
\langle T_{VV} \rangle&=&-\frac{1}{12\pi}C^{1/2}\partial^2_VC^{-1/2}
\n\\
&=&\frac{1}{24 \pi} \bigg[\frac{C_{,VV}}{C}-\frac{3}{2}\frac{(C_{,V})^2}{C^2} \bigg]\,,
\\
\langle T_{UV} \rangle&=&\frac{RC}{96\pi}=\frac{1}{24 \pi}\partial_U \partial_V \ln C\,, \label{TUV}
\ea
where $C$ is the conformal factor introduced in the above metric and $R$ is the scalar curvature.

Now let us also introduce a null coordinate $u$ affine on $ \mathscr{I}^+_{\rm right}$ such that
\be
U=p(u)\,;
\ee
from this we  get
\be
\partial_U=\dot{p}^{-1}\partial_u\, .
\ee
In terms of the set $(u,V)$, the metric reads
\be
ds^2=\bar C(u, V) dudV\,,
\ee
with 
\be
\bar C(u, V)=\dot p(u) C(U,V)\,.
\ee

Assuming that the observer is always outside the collapsing star, $\bar C(u, V)$ would be the metric component of a static spacetime. 
In terms of this newly defined null coordinate,  a simple computation shows that  $T_{UU}$ is given as
\be 
\langle T_{UU} \rangle=-\frac{\dot{p}^{-2}}{12\pi}\left[\bar C^{1/2}\partial^2_u\bar C^{-1/2}-\dot{p}^{1/2}\partial^2_u \dot{p}^{-1/2}\right]\,. \label{TUU}
\ee

Now $T_{VV}$ will have only a static contribution if $V=v$ but if the affine null coordinate  on $ \mathscr{I}^+_{\rm left}$ is defined as
\be
V=q(v)
\ee
and we define $C'(U,v)=\dot q(v) C(U,V)$, $T_{VV}$ is given as
\be 
\langle T_{VV} \rangle=-\frac{\dot{q}^{-2}}{12\pi}\left[ C'^{1/2}\partial^2_v C'^{-1/2}-\dot{q}^{1/2}\partial^2_v \dot{q}^{-1/2}\right] \label{TVV}\,.
\ee

As mentioned earlier  $\bar C(u, V)$ is the metric component of a static spacetime, so all the dynamics of the collapsing geometry is captured in the $\dot{p}$ term of (\ref{TUU}). In the above analysis, by using another affine null coordinate, we can differentiate between the static contribution to the RSET and the one due to the  dynamics associated with the collapse \cite{Barcelo:2007yk}.

\subsection{RSET for different vacuum states.}

Capturing the dependence at different radii of the RSET components would require a knowledge of the full $p(u)$ at any value of $u$, i.e. to specify a collapse history.
However, this would lead to the inclusion of transient effects which are not relevant for the present discussion. For this reason, we shall here rely on the fact that, well after the collapse has settle down, the black hole geometry is formally indistinguishable from that of an eternal configuration~\cite{Racz:1992bp,Racz:1995nh} (where the form of $p(u)$ is simply fixed by the geometry, see (\ref{p})). 

So, in order to extract physical information from the RSET, we shall compute the energy density and the flux experienced by an observer at constant Kruskal position long after the collapse has taken place in the two physically relevant states for Hawking radiation in the eternal black hole case, namely the Unruh and the Hartle--Hawking    states. We shall start in this Section by explicitly evaluating the general expressions for the RSET components expectation values.
 
Using (\ref{p}), we get the relations
\ba
\dot p(u)&\equiv& \partial_u p(u)= -\frac{p(u)}{2r_s}\, ,\\
\ddot p(u)&=&\frac{p(u)}{4r_s^2}= -\frac{\dot p(u)}{2r_s}\, .
\ea
For computing the first term of \eqref{TUU} we can write
\be\la{T1}
\bar C^{1/2}\partial_u^2 \bar C^{-1/2}= \frac{3}{4} \bar C^{-2} \left(\partial_u \bar C\right)^2-\frac{1}{2} \bar C^{-1}\partial_u^2 \bar C\,.
\ee
Using  the metric conformal factor $C$ from \eqref{kruskal} we get
\ba
\partial_u \bar C &=& \partial_u [\dot p(u) C]= \ddot p C +\dot p \partial_u C\n\\\n
&=& \dot p(u) \left( -  \frac{1}{2r_s}+  \frac{r^2-r_s^2}{2r^2 r_s} \right) C\\
&=& -\frac{r_s}{2r^2}  \bar{C}\,,
\ea
and
\ba
\partial_u^2 \bar C = -\frac{1}{2}r_s\partial_u\bigg( \frac{\bar{C}}{r^2}\bigg)
=\frac{r_s^2}{4r^4} \bar{C}-\frac{1}{2}\frac{r_sf(r)\bar C}{r^3}\, .
\ea

Using the above relation in (\ref{T1}) we have
\ba\la{22a}
&&\bar C^{1/2}\partial_u^2 \bar C^{-1/2} 
=\frac{3}{4}\bar{C}^{-2}\bigg[\frac{r_s^2}{4r^4}  \bar{C}^2 \bigg]
\n\\&-&\frac{1}{2}\bar{C}^{-1}\bigg[\frac{r_s^2}{4r^4} \bar{C}-\frac{1}{2}\frac{r_sf(r)\bar C}{r^3} \bigg]
\n\\&=&
-\frac{3}{16}\frac{r_s^2}{r^4}+\frac{r_s}{4r^3}
-\frac{3}{4}\frac{M^2 G ^2}{r^4}+\frac{M G }{2r^3},
\ea 
where $f(r)$ is given in \eqref{fr} and we used $r_s=2M G $ in the last step.
For the second term on the r.h.s. of \eqref{TUU}, we have
\be\la{22b}
\dot p^{1/2}\partial_u^2 \,\dot p^{-1/2}=-\frac{\dot p^{1/2}}{2}\partial_u \left(\frac{\ddot p}{\dot p^{3/2}}\right)
=\frac{1}{(8M G )^2}\,.
\ee
We are now ready to compute explicitly the expectation value of the different RSET components for the Hartle--Hawking ($|H\rangle$) and Unruh ($|U\rangle$) states. 

We can start by observing that for the $T_{UU}$ and $T_{UV}$ components, the expectation values are the same in the two vacuum states \cite{birrell1984quantum}. Therefore, in the following we simply denote
\ba
&&\langle T_{UU} \rangle\equiv \langle H| T_{UU}|H \rangle=\langle U| T_{UU}|U \rangle\,,\la{TUUdef}\\
&&\langle T_{UV} \rangle\equiv \langle H| T_{UV}|H \rangle=\langle U| T_{UV}|U \rangle\,. \la{TUVdef}
\ea
By means of (\ref{22a}), (\ref{22b}), $\langle T_{UU} \rangle$ is given by
\ba
\label{TUUfinal}
\langle T_{UU} \rangle &=&\frac{\dot{p}^{-2}}{24\pi}\bigg[\frac{3}{2}\frac{M^2 G ^2}{r^4}-\frac{M G }{r^3} +\frac{1}{32M^2 G ^2}\bigg] 
\n\\
&=&(768\pi M^2 G ^2)^{-1}\frac{V^2}{4r^2}e^{-r/M G }\n\\
&\cdot&\bigg[ 1+\frac{4M G }{r}+\frac{12M^2 G ^2}{r^2}\bigg]\,. 
\ea
  To compute $\langle T_{UV}\rangle$ we use (\ref{TUV}), from which
\ba
\langle T_{UV} \rangle&=&\frac{1}{24 \pi}\partial_U \partial_V \ln C=\frac{1}{24 \pi}(\dot{p}\dot{q})^{-1}\partial_u \partial_v \ln C
\n\\
&=&-\frac{1}{96 \pi}(\dot{p}\dot{q})^{-1}C\partial_r^2C.
\ea
Using $C(t,r)$ from (\ref{kruskal}) and the exact values of $q(u)$ and $p(v)$, we get
\be
\langle T_{UV}\rangle=-\frac{M^2 G ^2}{12 \pi r^4}e^{-r/2M G } \,. \label{TUVfinal}
\ee

On the other hand, the dependence  of $\langle T_{VV} \rangle $ on the state in which we are computing the expectation value is important. 
For the Hartle--Hawking state (eternal black hole scenario, non-singular vacuum state in both past and future horizons) in Kruskal coordinates the modes are given by $e^{-i\omega U},e^{-i\omega V}$, where we defined $V$ as
\be
V\equiv q(v)=2r_se^{v/2r_s}\,.
\ee

Using this definition of $V$ we can proceed in a similar way as for the computation of  $\langle T_{UU}\rangle$. From (\ref{TVV}),  we obtain
\ba\la{TVVH}
\langle H| T_{VV}|H\rangle\!\!&=&\!\frac{\dot{q}^{-2}}{24\pi}\bigg[\frac{3}{2}\frac{M G ^2}{r^4}-\frac{M G }{r^3} +\frac{1}{32M G ^2}\bigg] 
\n\\
\!&\!=\!&\!(768\pi M^2 G ^2)^{-1}\frac{U^2}{4r^2}e^{-\frac{r}{M G }}\n\\
&\cdot& \bigg[ 1+\frac{4M G }{r}+\frac{12M^2 G ^2}{r^2}\bigg]\,.
\ea

For the Unruh state in Kruskal coordinates,  the modes are given by $e^{-i\omega U},e^{-i\omega v}$ and there is no regularization condition imposed in the past horizon. The expectation value of the $T_{VV}$ component can be 
obtained from the relation
\ba
\langle U| T_{VV} |U\rangle=16M G ^2 \dot{q}^{-2}\langle U| T_{vv} |U\rangle \,, \label{Tvvu}
\ea
where $\langle U| T_{vv} |U\rangle$ can be computed from
\ba
\langle U|T_{vv}|U \rangle=-\frac{1}{12\pi}f(r)^{1/2}\partial^2_vf(r)^{-1/2}\,
\ea
using $f(r)=\big(1-\frac{2M G }{r} \big)$, as follows from the metric of a black hole  in static Schwarzschild coordinates. We have
\ba
\langle U|T_{vv} |U\rangle=\frac{1}{24\pi}\bigg[ \frac{3M^2 G ^2}{2r^4}-\frac{M G }{r^3}\bigg]\,,
\ea 
and from (\ref{Tvvu}) we get
\be
\langle U| T_{VV} | U\rangle=\frac{1}{6\pi}\frac{M^2 G ^2}{V^2}\bigg[ \frac{3M^2 G ^2}{2r^4}-\frac{M G }{r^3}\bigg] \,. \label{TVVu}
\ee

\subsection{Energy density }

We now have all the ingredients to extract physical information from the RSET. Let us first analyze the energy density as measured in the frame of an observer moving along fixed position in Kruskal coordinates.

Let us consider an observer at a given Kruskal position with  2-velocity $v^\mu=C^{-1/2}(1,0)$ (in $[T,X]$ coordinates)\footnote{ This choice of trajectory is not geodesic; however the acceleration that the observer experiences is irrelevant compared to the Hawking temperature and one can show easily that the acceleration vanishes at the horizon. One might think that a free falling observer would have been a better choice. However, the problem with such choice would be  the non-zero radial velocity of the free falling observer at the horizon, as well as near the horizon. In that case, it would then be difficult to separate out the Hawking radiation contribution from  other kinematical effects \cite{Barbado:2011dx} }. The energy density, $\rho$,  measured by this observer for the Unruh state is given by
\ba\la{rho}
\rho=\langle U| T_{\mu\nu} |U\rangle v^\mu v^\nu=C^{-1}\langle U |T_{TT} |U\rangle 
\n\\
=C^{-1}\langle U |T_{VV}+ T_{UU}+2T_{UV}|U\rangle \,.
\ea
 Using (\ref{TUUfinal}),  (\ref{TUVfinal}), (\ref{TVVu}) we can compute the energy density exactly and we plot it in FIG. \ref{ed} (where $\alpha\equiv r/r_s$).
\begin{figure} [htb]
\includegraphics[scale=.65]{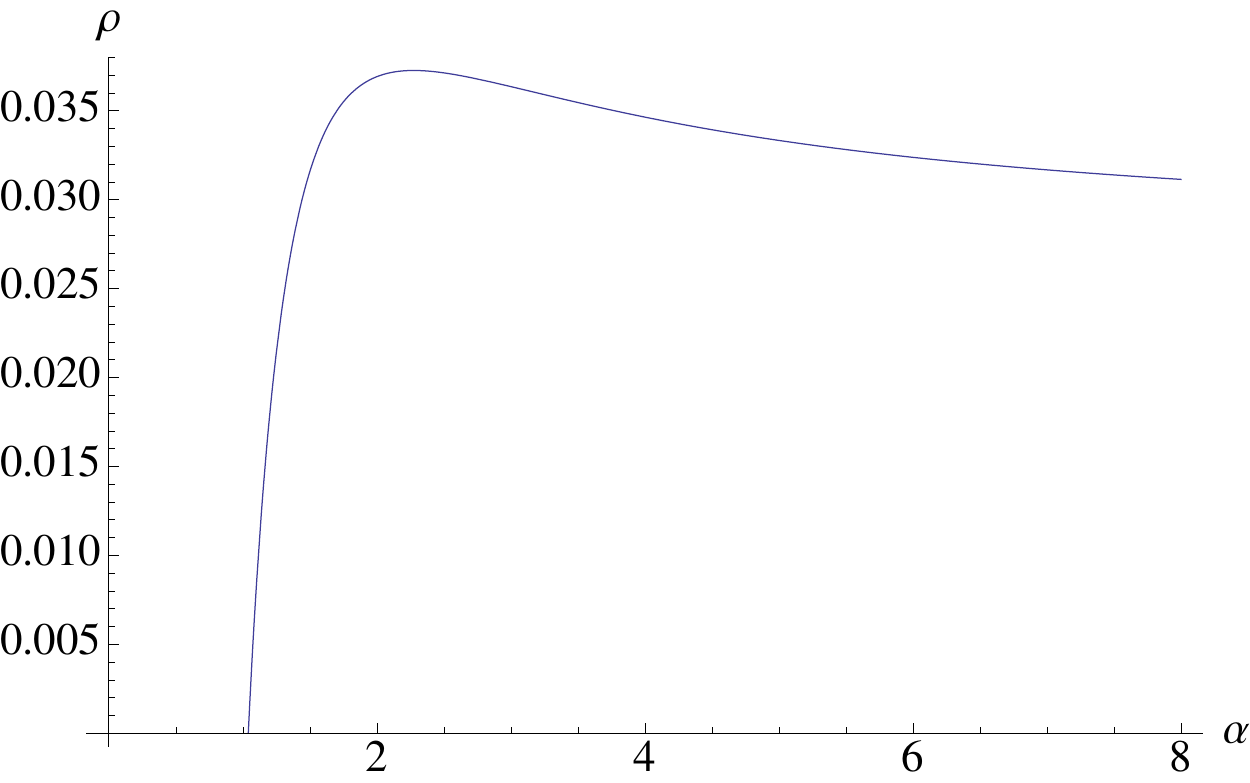}
\caption  {Plot of the energy density at a given time as a function of the the radial distance from the centre of the black hole in Unruh state at a given instant of time.}\la{ed}
\end{figure}
 
The energy density \eqref{rho} blows up at the horizon $(r=2M)$ since we are computing the energy density as observed by a free falling (in Kruskal coordinates) observer in the Unruh state 
which is well known to be ill defined on the past horizon. {Such divergence arises from the $1/V^2$ term in the component \eqref{TVVu} when $V=0$, i.e. at the past horizon. The horizon location condition in Schwarzschild radial coordinate, $\alpha=1$,  cannot distinguish between past and future horizons and thus the divergent contribution would enter in the plot above of the energy density expression \eqref{rho} when evaluated at $\alpha=1$. However, a free falling observer at the future horizon would not see this divergence, which is just an artifact of Kruskal coordinates \footnote{{ Let us stress that also the calculation in~\cite{Barcelo:2007yk} of the RSET components in the collapse scenario shows that at the white hole horizon the Unruh state will necessarily be singular. This can be easily realised by applying time reversal to the subdominant terms in the dynamical contribution \eqref{22b} derived in~\cite{Barcelo:2007yk} (see Eq. (52) there), which then shows an exponentially growing flux at the white horizon which very rapidly would create a divergence in the $T_{UU}$ component of the RSET soon after horizon formation.}}.
This is a well known fact already pointed in \cite{Unruh:1977ga}. For this reason, we have removed the point $\alpha=1$ in the plot shown in FIG. \ref{ed}. 

Near the horizon the energy density becomes negative; these negative values are attained closer to the horizon as the energy density is measured at later times. We show this near horizon behavior in the two plots in FIG. \ref{ed2}, where the first is evaluated at the same time as the plot in FIG. \ref{ed} and the second one at a close instant after (a similar behavior was found also in \cite{Eune:2014eka}); the negative divergent behavior of the energy density at the horizon is clear from the plots. 

However, let us remark again that this divergence is just fictitious for an observer crossing the future horizon $U=0$ at a given value of $V>0$ and it is an inevitable feature of plotting the energy density in the Unruh state as a function of $r$ for a fixed instant of time $t$.  

One way to avoid this misleading behavior of the energy density plot at the horizon could be to show it as a function of $U$ for given $V= const >0$; this would indeed remove the singularity from the plot since the point $\alpha=1$ would now correspond to $U=0$, i.e. to the future horizon where the Unruh state is regular. However, from such plot it would be very
difficult to extrapolate the information about how the energy density is distributed in the $r$ coordinate for {\it fixed} time $t$, since fixing $V$ and letting $U$ run imply that different values of $U$ correspond to different values of $r$ {\it and} $t$. 
\begin{figure} [htb]
\centering
 \subfigure
        \centering
        \includegraphics[height=1.6in]{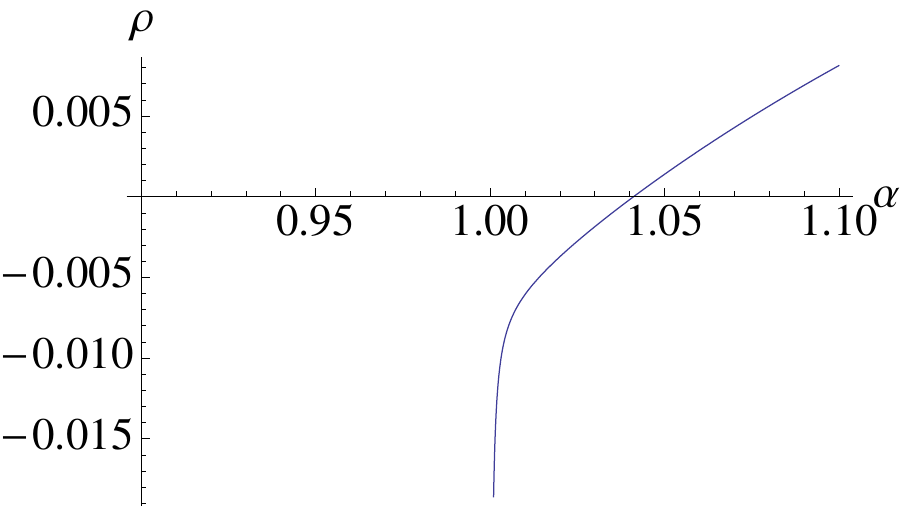}
    ~ 
    \subfigure
    \centering
    \includegraphics[height=1.6in]{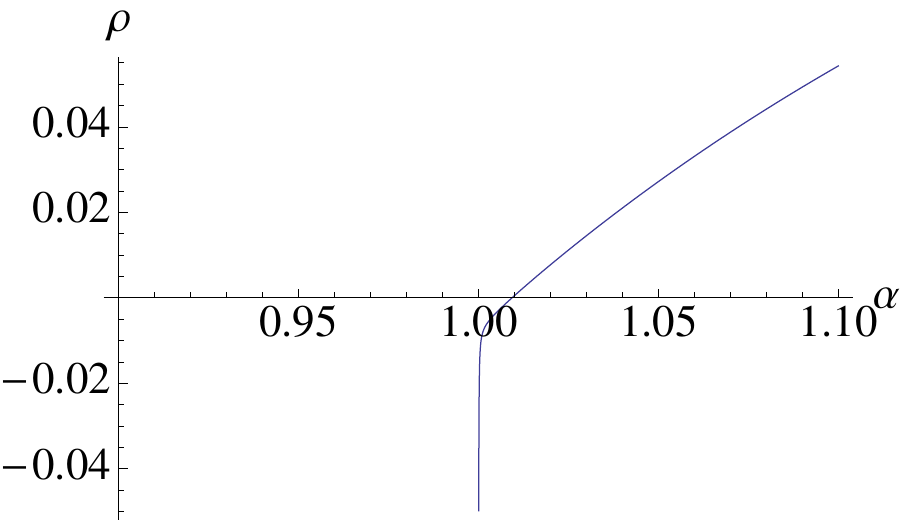}
    \caption{Near horizon behavior of the energy density in the Unruh state at different times. The first plot corresponds to the same instant of time as the plot in FIG. \ref{ed}; the second one to a close instant after.}
\la{ed2}
\end{figure}
}

The significant aspect of the plot in FIG. \ref{ed}  for us is the peak in the distribution of $\rho$ that is obtained outside the horizon which is at $r \approx 4.32 MG$. Quite in agreement with our heuristic prediction based on the gravitational analogue of the Schwinger effect. { Let us point out that, although we have shown the plot at a given instant of Killing time, the behavior of the energy density remains the same at any time, in particular the presence of the peak at the same location persists; the only difference is that the value of the energy density increases since it accumulates, given that we are not taking into account the effect of back-reaction. }
 
 To get a non-singular energy density plot for the free falling observer we should consider  the Hartle--Hawking state. 
This is given by
\ba\la{rhoH}
\rho&=&\langle H| T_{\mu\nu} |H\rangle v^\mu v^\nu=C^{-1}\langle H |T_{TT} |U\rangle 
\n\\
&=&C^{-1}\langle H |T_{VV}+ T_{UU}+2T_{UV}|H\rangle\,.
\ea
Using the expectation values given in  (\ref{TUUfinal}), (\ref{TUVfinal}), (\ref{TVVH}), we can plot the energy density \eqref{rhoH}  with respect to radial distance parametrized by $\alpha$. This is shown in FIG. \ref{edH}, where we see a similar nature of the distribution with a peak outside the horizon; however, as expected, in this case the energy density is regular everywhere.  Remarkably, the peak is located at $r\approx 4.37MG$, in close agreement with our heuristic findings.
\begin{figure} [htb]
\includegraphics[scale=.65]{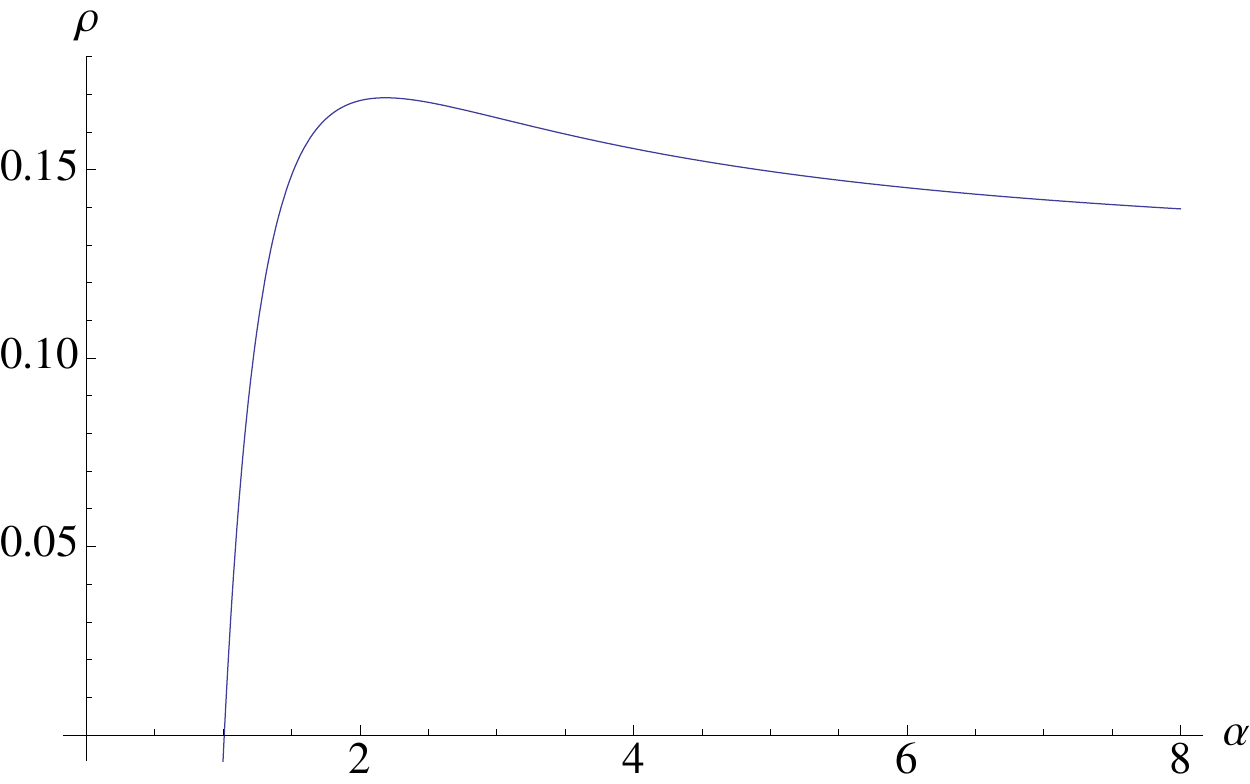}
\caption  {Plot of the variation of energy density computed in Hartle--Hawking state with respect to the radial distance from the centre of the black hole at fixed time measured in the static frame. Notice that close to the horizon the energy density is negative also in this case, { but it remains finite at the horizon due to the non-divergent behavior of the $T_{VV}$ component \eqref{TVVH} in the Hartle--Hawking vacuum}.}\la{edH}
\end{figure}

 These results strongly support our previous claim that the radiation density is maximized in a region outside the horizon. 
We now show that a similar behavior with a peak away from the horizon is exhibited also by the flux part of the RSET.

\subsection{Flux}
The flux of the Hawking radiation in the Unruh vacuum is given by \cite{Ford:1993bw} \footnote{In the Hartle--Hawking vacuum the flux vanishes due to the thermal equilibrium of the state.}
\ba
F=-\langle U|T_{\mu\nu}|U\rangle v^\mu z^\nu\,,
\ea
where $v^\mu$ is the velocity of the observer and $z^\nu$ is the contravariant component of the normal to the observer. Let us consider a static observer at fixed distance in a Kruskal frame with $v^\mu=C^{-1/2}[1,0]$ and indicate the normal vector as $z^\nu=[A,B]$. The latter has to satisfy  the following conditions
\ba
g_{\mu\nu}z^\mu z^\nu=-1, \  z^\mu v_\mu=0\,.
\ea
Using the second relation we get $A=0$ and from the first relation we get $B=C^{-1/2}$.
Therefore, 
$z^\nu=C^{-1/2}[0,1]$. 

 Using these expressions for $v^\mu, z^\nu$, we get
\ba
F=-C^{-1}\langle U| T_{TX}|U\rangle=C^{-1}\langle U|[-T_{VV}+T_{UU}]|U\rangle\,.
\ea

Plugging in the expectation values \eqref{TUUfinal}, \eqref{TVVu}  found above, we can plot the flux as a function of $\alpha$. This is shown in FIG. \ref{fig:flux}. { Also in this case the plot of the flux would receive a fictitious (for a free falling observer at the future horizon) divergent contribution from the component \eqref{TVVu}, and we have thus removed the point $\alpha=1$ from the plot, thus avoiding the divergence at the past horizon $V=0$.
}
\begin{figure}[htb]
\scalebox{0.65}{{\includegraphics{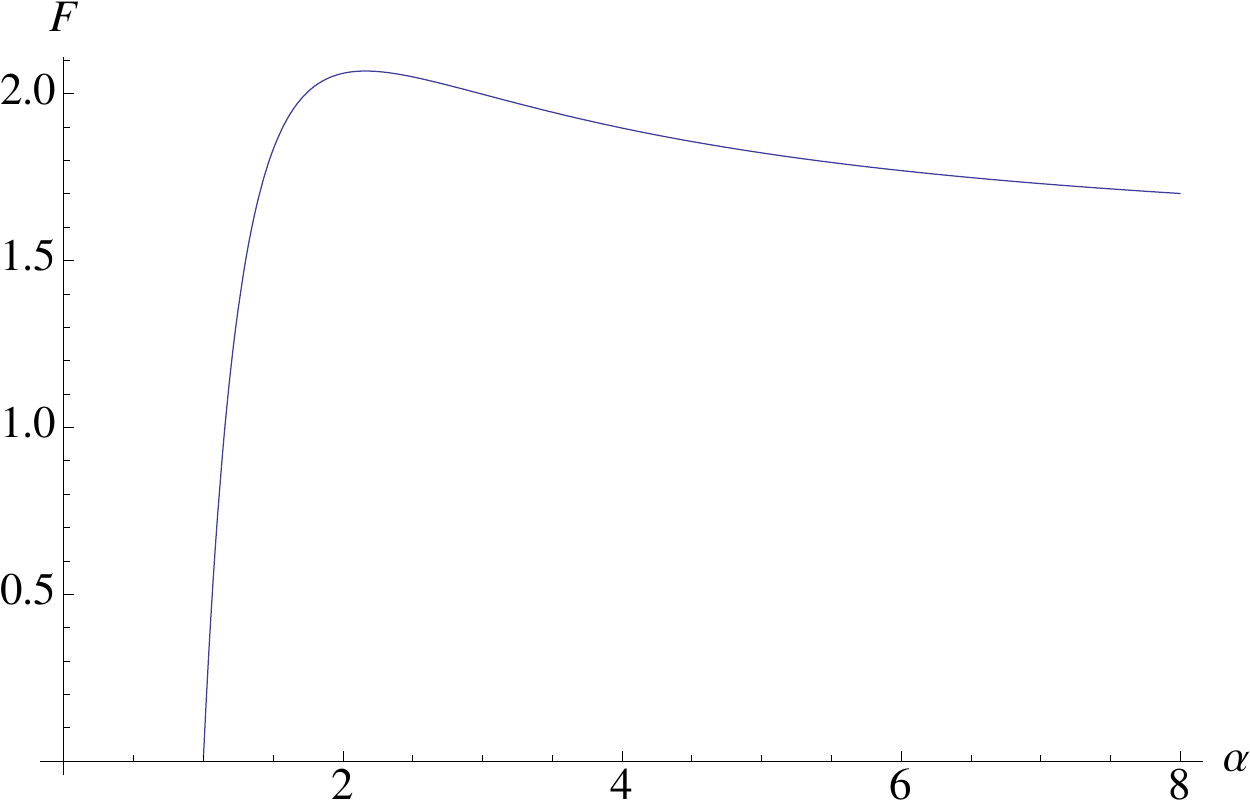}}}
\caption {This plot shows the variation of the flux of Hawking radiation with respect to the radial distance as measured by an observer in the Unruh state at a given instant of time.}\la{fig:flux}
\end{figure}
We see that the flux has a maximum at $r=4.32MG$ and most of the contribution to the Hawking radiation comes from a region between the horizon and $r\approx 6MG$.

 { Let us remark that our findings are in line with the analysis of the 2-dimensional RSET done in \cite{Giddings:2015uzr} where it was shown that the ingoing and outgoing null components of the stress tensor would build up to their asymptotic values in a region outside the horizon. In our analysis we have been more precise in confirming this result by choosing an observer and explicitly computing the values of the energy density and the flux outside the horizon as measured by the observer.   }


\section{Summary and Discussion}
It has been widely believed that Hawking radiation originates from the excitations close to the horizon and this eventually suggested some drastic modification of the states in the near horizon regime  as a resolution to the information loss paradox \cite{Almheiri:2012rt,Papadodimas:2012aq,Braunstein:2009my,Almheiri:2013hfa}. One of the primary reasons for such an argument is based on the way Hawking did his original calculation, tracing back the modes all the way from future infinity to the past null infinity through the collapsing matter so that one has a vacuum state at the horizon for a free-falling observers. 

The other disturbing feature about this argument is, when the modes are traced back they become highly blueshifted near the horizon and we are not well aware of the laws of physics in such high transplanckian domain. Some resolutions to the above problem has been proposed several time in the literature \cite{PhysRevD.51.2827,Corley:1996ar,PhysRevD.44.1731} but they all demand some challenging modification to our present knowledge of gravitation or quantum field theory. 

{Let us stress, however, that the UV departures from Lorentz invariance through the introduction of  a fundamental cutoff postulated in \cite{Unruh:1994je, Jacobson:1991gr} are relevant only very close to the horizon for large black holes (in units of the Lorentz breaking scale). Hence, even contemplating such scenario, our analysis in section \ref{SET} would be basically unchanged and unaffected away from the horizon, as also stressed in the similar analysis carried out in \cite{Brout:1995wp}.

 In this paper we have shown evidence that the Hawking quanta originate from a region which is far outside the horizon, which can be called a black hole {\it atmosphere}. More precisely, from the plots of the energy density and the flux in the Unruh state we get a maximum at $r\approx 4.32MG$, for the energy density in the Hartle--Hawking state the peak is at $r\approx 4.37MG$. This is strikingly close to our previous finding for an origin at about $r\approx 4.38MG$ for the peak of the thermal spectrum using the heuristic argument based on tidal forces. By large this is also in agreement with some previous claims using various other methods, such as calculating the effective radius of a radiating  body using the Stefan--Boltzmann law or computing the effective Tolman temperature \cite{Giddings:2015uzr,Hod:2016hdd,Eune:2015xvx, Kim:2016iyf}, { as well as  in close correspondence with the results of the
study of the null component of the stress-energy tensor in the Unruh vacuum
of \cite{Parentani:1992me}. }
 
{  Given the presence of a quantum atmosphere where the Hawking quanta are generated and which extends well beyond the black hole horizon, as originally suggested in \cite{Giddings:2015uzr}, it would be interesting to investigate how its effective radius is affected by going to higher dimensions. Applying the  Stefan-Boltzmann radiation law argument proposed in \cite{Giddings:2015uzr} for the  (3 + 1)-dimensional case to $(D + 1)$-dimensional
Schwarzschild black holes, it was found in \cite{Hod:2016hdd} that the effective radius gets squeezed towards the black hole horizon as the number of spatial dimensions increases. 

Given that there is no derivation of the RSET components in dimensions higher than (1+1), we cannot apply the argument presented in Section \ref{SET} to confirm this result.
However, the heuristic derivation that we presented in Section \ref{Schwinger} could be easily generalized for any arbitrary number of dimensions. Without presenting a complete derivation, we can understand in a qualitative way how the quantum atmosphere can be effected by going to arbitrary $(D+1)$ higher dimensions  by  considering the fact that the Hawking temperature  scales as $T_{BH}=\frac{(D-2)\hbar}{8\pi MG}$, where $D$ is the number of spatial dimensions. It can be shown that this dimensional scaling of the temperature, along with the modification of the Schwarzschild metric for an arbitrary $D$, would yield, for given $r=r_*$, a higher value of $\omega_r$ \eqref{omegarini} as $D$ increases. At the same time, it can be shown from dimensional arguments that the work done by the tidal force must decrease in value for the same given $r$ as $D$ increases. This implies that, for a fixed $D>3$, the peak of the Hawking radiation spectrum corresponds to an higher value of energy than in the $D=3$ case and, in order for the gravitational field to be able to provide enough work to reach such amount of energy, the outgoing partners comprising the bulk of the spectrum at infinity must go on-shell closer to the horizon. Our Schwinger effect argument thus confirms in a qualitative way the relation obtained in \cite{Hod:2016hdd} for the decrease of the effective radius in the regime $D\gg 1$.}
 
 If the radiation has a long distance origin then we might not need to worry about the transplanckian issue at the horizon. Moreover, concerning the fundamental issue of unitarity of black hole evaporation, this result suggests to consider some effect operational at this new scale  in order to eventually restore unitarity of Hawking radiation .  A possible scenario is the one of non-violent nonlocality advocated in \cite{Giddings:2012bm, Giddings:2012gc}; see also the proposal of \cite{Nomura:2014woa, Nomura:2014voa}. We hope that the present contribution will stimulate further investigations in these directions.

\section{Acknowledgment}
 We thank Renaud Parentani, Sebastiano Sonego and Matt Visser for illuminating discussions. We also acknowledge the John Templeton Foundation for the supporting grant \#51876.

\appendix 
\section { Kruskal frame.}
 We want to examine the components of the RSET in a globally well defined coordinate system free of any pathological behavior (other than a true curvature singularity, like in the center of a black hole). For this purpose the Kruskal coordinate frame is an appropriate choice. The Kruskal metric is given as

\be
ds^2= \frac{ r_s}{r}e^{-r/r_s}dUdV \,,\label{kruskal}
\ee
where $r_s$ is the radius of the event horizon. For this coordinate system we have
\ba
&&U=p(u)=-2r_s e^{-u/2r_s}\,,  \label{p}\\
&&V=q(v)=2r_s e^{v/2r_s}.
\ea

The affine null coordinate $u,v$ in terms of radial distance from the centre of the black hole, $``r"$, and time, $``t"$, as measured by a static observer is given as 
\ba
u=t-r_{*}= t- \left[r+r_s\ln{\left(\frac{r}{r_s}-1\right)}\right],
\\
v=t+r_{*}= t+ \left[r+r_s\ln{\left(\frac{r}{r_s}-1\right)}\right].
\ea
also 
\ba
&&\partial_u=\frac{\partial r_{*}}{\partial u}\partial_{r_{*}}=-\frac{1}{2}\partial_{r_{*}}= -\frac{1}{2}f(r)\partial_{r}\,, \la{partialu}
\\
&&\partial_v=\frac{\partial r_{*}}{\partial v}\partial_{r_{*}}=\frac{1}{2}\partial_{r_{*}}=\frac{1}{2}f(r)\partial_r.
\ea
where we used
\be \la{fr}
\frac{dr_{*}}{dr}=[f(r)]^{-1}=\bigg(1-\frac{r_s}{r}\bigg)^{-1}.
\ee
We can also define a set of time like and radial coordinates $(T,X)$ as
\ba
T=\frac{1}{2}(V+U),X=\frac{1}{2}(V-U).
\ea
Using this metric (\ref{kruskal}) is given as
\be
ds^2= \frac{ r_s}{r}e^{-r/r_s}(dT^2-dX^2)\,.
\ee

\bibliographystyle{apsrev-title}
\bibliography{hawking-quanta.bib}

\end{document}